# Enhancing Portfolio Optimization with Deep Learning Insights

Brandon Luo
Georgia Institute of Technology
`zluo309@gatech.edu`

Jim Skufca
Georgia Institute of Technology
`jskufca3@gatech.edu`

## Abstract

*Our work focuses on deep learning (DL) portfolio optimization, tackling challenges in long-only, multi-asset strategies across market cycles. We propose training models with limited regime data using pre-training techniques and leveraging transformer architectures for state variable inclusion. Evaluating our approach against traditional methods shows promising results, demonstrating our models' resilience in volatile markets. These findings emphasize the evolving landscape of DL-driven portfolio optimization, stressing the need for adaptive strategies to navigate dynamic market conditions and improve predictive accuracy.*

## 1. Introduction

Modern portfolio theory (MPT) and mean-variance optimization (MV) are pivotal in capital management, offering many benefits to society. They enable risk reduction through diversified asset allocation, guide efficient capital use for productive investments, empower informed decision-making for investors, and contribute to market stability by discouraging excessive speculation. The integration of MPT and MV not only shapes capital management practices but also fosters economic resilience and financial progress.

The history of portfolio theory traces back to seminal work by Harry Markowitz, whose pioneering efforts laid the groundwork for MPT. Subsequent contributions by researchers such as Eugene F. Fama, Kenneth French, William Sharpe, and Stephen Ross further expanded the theoretical underpinnings of portfolio management. Despite the significant progress made, critiques over time have highlighted certain limitations inherent in these traditional approaches. However, recent advancements in computational capabilities, coupled with the availability of open-source tools, have ushered in a new era of innovation in machine learning (ML) algorithms.

Our research explores ML applied to portfolio optimization, building upon traditional portfolio theory while recognizing its limitations. During our literature review, we discovered a paper by researchers at Oxford University (Oxford) that employs DL techniques, showing promising initial outcomes [1]. Our paper aims to contribute to this field by confirming and evaluating the methodologies from the Oxford study, testing the robustness of their approach, and expanding upon their work by exploring different feature spaces, neural network architectures, and loss functions, seeking potential enhancements.

Our work is meaningful in a new field where solutions are still developing. Portfolio optimization, especially with ML, is complex and evolving, showing uncertainty and diverse approaches. Advancing here matters for investors, institutions, and wider societal impacts like stable economies and smart capital use.

Our paper is structured as follows: we review background and related work in portfolio optimization, including MPT and DL techniques. The motivation section emphasizes long-only, multi-asset integrated portfolio optimization. We detail our approach, methods, and data sources, followed by presenting results, discussing findings, and offering insights for future research. In this report, all figures, except for Table 1, and the Work Division section are placed at the back of the document.

## 2. Background and Related Work

In the domain of portfolio optimization, extensive research has been conducted across two key areas: the established principles and constraints of MPT and the emergent exploration of DL techniques for portfolio optimization.

MPT, introduced by Markowitz in 1952, revolutionized asset management by integrating risk and return considerations [2]. Other pivotal models and theories include Arbitrage Pricing Theory (APT) by Stephen Ross [3], Fama and French Models [4], and William Sharpe's Sharpe ratio [5].

Researchers have addressed MPT's limitations like normality assumptions and constant correlations [6, 8]. They have developed advanced optimization techniques, merging Markowitz's efficient frontier with Sharpe's metric for comprehensive portfolio management [9,10], which underscores the crucial understanding of asset class correlations for diversification benefits [6, 8, 10].

ML researchers in portfolio optimization have explored a wide array of methodologies, including feature space analysis encompassing stock prices, returns, technical indicators [11, 12], textual data from financial news and reports [13-15], structured data from financial markets [16], and knowledge graphs [17]. Techniques like CNNs have been used for dimensionality reduction and insights extraction [16, 18, 19], while RNNs and LSTMs are favored for capturing temporal dependencies and nonlinear relationships in financial time series [19, 20]. Additionally, targeting in these studies has expanded to include stock and forex prices and returns [21, 22], with RL methods gaining traction for dynamic optimization and adaptive portfolio strategies [21, 23, 24].

Researchers have explored non-integrated and integrated methods. Non-integrated approaches break down optimization into separate components such as return, volatility, and correlation within an end-to-end framework [25, 26], while integrated approaches use learning algorithms directly on feature space and target variables, optimizing asset weights for risk-adjusted returns [20, 27].

Our review of the literature in ML portfolio optimization reveals a prevalent issue of overfitting to specific datasets among many studies, highlighting a lack of robust and transferable approaches. Consequently, significant limitations persist in current methodologies despite the extensive work in this field.

## 3. Approach

Our work builds upon the Oxford study's approach in long-only, multi-asset integrated optimization, addressing overfitting concerns. Their model fills a gap by focusing on diversified asset allocation for substantial wealth management. We aimed to assess and enhance their LSTM model's robustness, validate findings, and identify improvements.

### 3.1 Replicating and verifying previous work

#### 3.1.1 Methods

The Oxford approach rests on three core principles: the temporal nature of asset prices [28], the suitability of LSTM for feature learning and prediction, and the use of a Sharpe ratio-based loss function to optimize asset weights for maximizing risk-adjusted returns in a multi-asset portfolio. The Sharpe ratio is calculated as:

$$Sharpe\ Ratio = (R_p - R_f)/\sigma_p$$

where $R_p$ is the portfolio's expected return, $R_f$ is the risk-free rate of return, and $\sigma_p$ is the portfolio's standard deviation of returns.

The optimization strategy maximizes the Sharpe ratio via a neural network, bypassing traditional forecasting by linking asset returns and prices over a trading period. It ensures positive weights sum to one using a SoftMax function. Gradient ascent optimizes the framework during training, updating parameters iteratively for Sharpe ratio convergence [17, 24]. Oxford's LSTM architecture includes layers for input, LSTM, and SoftMax output, with mini-batches spanning $n$ days and using observations from the past $m$ days to optimize daily asset weights for the Sharpe ratio. Features are extracted from an $n$-day look-back window, linked to the next day's asset returns [1, 26].

During training, the Oxford team used a validation set comprising 10% of the training data. They identified the most effective model with hyperparameter tuning of LSTM with 64 units, a 50-day lookback window, and a mini-batch size of 64, achieving convergence within 100 epochs. During training, we found that a learning rate of 0.001 led to optimal model convergence and performance during training. We assumed a zero risk-free rate, aligning with the Oxford approach.

The Oxford team's training process spanned from 2006 to the end of 2010, with periodic retraining every two years using all available data up to that point to update parameters. Their testing period extended from 2011 to the end of April 2020, providing a comprehensive assessment of their model's performance over a significant timeframe.

#### 3.1.2 Analysis

In our study, we adopted the approach designed by the Oxford team, which involved analyzing the optimized portfolio generated by our LSTM model (LSTM) using a range of portfolio metrics. These metrics encompassed key aspects such as return (cumulative and annual returns), risk (volatility, max drawdown, downside deviation), profit (percent positive returns and average profit/average loss), and risk-adjusted return (Sharpe, Sortino).

To fully evaluate the performance of their optimized portfolio, the Oxford team compared its metrics with those of several benchmark portfolios. These benchmarks included mean variance optimized portfolios, maximum diversification portfolios, diversity-weighted portfolios, and fixed weight portfolios with varying allocation strategies.

We improved analysis robustness by evaluating portfolio Sharpe ratios rigorously, using the Mann-Whitney U test (MW) to handle autocorrelation in rolling Sharpe ratios and z-tests for comparisons across independent samples. We also refined our choice of benchmark portfolios. We kept the mean variance optimized benchmark (MVO) but dynamically optimized its weights (floored at 0.1 and capped at 0.9) using a three-year lookback period with quarterly resets. We also introduced a balanced portfolio (Balanced) with equal weights as a benchmark to address historical prediction biases seen in fixed weight portfolios used in previous studies, including those at Oxford.

#### 3.1.3 Data

The data used in our study mirrored that of the Oxford team, focusing on a four-asset portfolio chosen for their high trading volume and low return correlations. This portfolio included a broad stock portfolio (VTI), a broad bond portfolio (AGG), a commodities portfolio (DBC), and a volatility measure (VIX). Our dataset consisted of daily price and return data for these assets, totaling eight features. The time series spanned from 02/07/2006 to 04/30/2020, providing a substantial historical dataset for analysis. It is important to note that the data was unnormalized and downloaded directly from Yahoo Finance's API. Notably, we substituted the absence of an ETF for VIX with the daily closing values of the CBOE Volatility Index (^VIX) in our dataset. There were no missing values nor outliers in our dataset

| | Experiments | | | | | | | | | | | | | | | | |
|---|---|---|---|---|---|---|---|---|---|---|---|---|---|---|---|---|---|
| | Verification | | | | Time | | | Asset | | | Features | | | Alternative Design & Data | | | |
| Measure | LSTM | Oxford | MVO | Balanced | LSTM | MVO | Balanced | LSTM | MVO | Balanced | LSTM | MVO | Balanced | Transformer | LSTM | MVO | Balanced |
| Cumulative Return | 15.705 | 11.632 | 3.266 | 5.614 | 11.237 | 5.828 | 10.632 | 0.733 | 0.408 | 0.985 | 9.906 | 5.828 | 10.632 | 6.497 | 11.237 | 5.828 | 10.632 |
| Annual Return | 0.353 | 0.313 | 0.174 | 0.279 | 0.219 | 0.165 | 0.258 | 0.045 | 0.027 | 0.056 | 0.208 | 0.165 | 0.258 | 0.168 | 0.219 | 0.165 | 0.258 |
| Annual Volatility | 0.163 | 0.168 | 0.097 | 0.297 | 0.096 | 0.094 | 0.287 | 0.054 | 0.025 | 0.061 | 0.096 | 0.094 | 0.287 | 0.061 | 0.096 | 0.094 | 0.287 |
| Sharpe Ratio | 1.909 | 1.858 | 1.658 | 0.827 | 2.061 | 1.622 | 0.800 | 0.809 | 1.084 | 0.891 | 1.965 | 1.622 | 0.800 | 2.740 | 2.061 | 1.622 | 0.800 |
| Downside Deviation | 0.082 | 0.099 | 0.061 | 0.172 | 0.054 | 0.059 | 0.166 | 0.040 | 0.018 | 0.045 | 0.059 | 0.059 | 0.166 | 0.054 | 0.054 | 0.059 | 0.166 |
| Sortino | 3.728 | 3.135 | 2.649 | 1.430 | 3.644 | 2.591 | 1.377 | 1.087 | 1.492 | 1.228 | 3.186 | 2.591 | 1.377 | 3.138 | 3.644 | 2.591 | 1.377 |
| Max Drawdown | 0.111 | 0.102 | 0.153 | 0.178 | 0.114 | 0.153 | 0.178 | 0.204 | 0.059 | 0.142 | 0.172 | 0.153 | 0.178 | 0.106 | 0.114 | 0.153 | 0.178 |
| % of + Return | 0.554 | 0.537 | 0.514 | 0.477 | 0.553 | 0.512 | 0.469 | 0.558 | 0.552 | 0.544 | 0.547 | 0.512 | 0.469 | 0.575 | 0.553 | 0.512 | 0.469 |
| Ave P/Ave L | 1.318 | 1.518 | 1.327 | 1.284 | 1.281 | 1.312 | 1.313 | 0.926 | 0.988 | 0.985 | 1.238 | 1.312 | 1.313 | 1.256 | 1.281 | 1.312 | 1.313 |

Table 1. Portfolio performance for various strategies and optimization techniques.

### 3.2 Testing approach robustness

We conducted three experiments primarily to assess the robustness of their approach. These experiments included an extended time-period, a different asset composition, and expanded feature space.

#### 3.2.1 Time experiment

The time experiment is notable because the Oxford data concluded before the COVID-19 (or pandemic). Extending the time-period into the pandemic, with its intense market volatility and daily reactions to unprecedented events, prompted us to investigate how well their approach would perform under these extreme circumstances.

In this experiment, we adopted the same LSTM architecture and hyperparameter settings as the Oxford study. However, it is important to note that everything needed to be recomputed to accommodate the extended time-period through 12/31/2023. This included re-evaluating benchmark portfolios, statistical tests, portfolio metrics, and benchmarks, ensuring a comprehensive analysis over a longer timeframe that encompassed significant market events and economic shifts.

#### 3.2.2 Asset experiment

After observing the impressive results achieved by the Oxford team with their 4-asset portfolio, our second experiment explored the transferability of their approach across different asset classes.

We maintained continuity in our second experiment by using the same architecture and hyperparameters as the Oxford study, but we extended the time-period to the end of 2023. We followed Oxford's guidance by selecting assets with low or negative correlation, choosing the Wilshire Total Market Index (Wilshire) and ICE BofA US Corporate Index (BofA) for stocks and bonds, respectively. Additionally, we substituted spot gold for the commodity fund and replaced VIX with Fed-funds, a cash alternative asset (Figures 1, 2). Three of these instruments' data were sourced from the Federal Reserve Banks' website (fred.stlouisfed.org), while gold prices were obtained from Kaggle (kaggle.com)

| Symbol | Description | Source | Symbol | Description | Source |
|---|---|---|---|---|---|
| *Assets* | | | *Economic Indicators* | | |
| VTI | US total stock index | Yahoo | HOUST5F | Housing Starts | FRED |
| AGG | US aggregate bond index | Yahoo | PERMIT | Building Permits | FRED |
| DBC | US commodity index | Yahoo | GDP | GDP | FRED |
| ^VIX | CBOE Volatility Index | Yahoo | CPIAUCSL | CPI | FRED |
| WILL5000IND | Total markets index US | FRED | PPIACO | PPI | FRED |
| BAMLCC0A0CMTRIV | ICE BofA US Corporate Index | FRED | DIVIDEND | Corporate Profits | FRED |
| FEDFUNDS | Federal funds rate | FRED | INDPRO | Industrial Production | FRED |
| --- | Spot gold | Kaggle | HSN1F | New Home Sales | FRED |
| | | | UNDCONTSA | Homes Under Construction | FRED |
| *Financial Markets* | | | IMPGSCA | Imports Total | FRED |
| T10Y2Y | 10/2 Treasury Bond Spread | FRED | PCE | Personal Consumption Expeditures | FRED |
| FEDFUNDS | Fed Funds Rate | FRED | ICSA | Unemployment Claims | FRED |
| | | | CSCICP03USM665S | Consumer Sentiment | FRED |
| *Economic Indicators* | | | USALOLITONOSTSAM | Leading Indicators | FRED |
| BSCICP03USM665S | Business Confidence | FRED | CSCICP03EZM665S | Euro Area Confidence | FRED |
| UNRATE | Unemployment Rate | FRED | RECPROUSM156N | Recession Probability | FRED |
| TCU | Capacity Utilization | FRED | | | |

Figure 1 - Data table

| Correlation Table | | | | |
|---|---|---|---|---|
| | VTI | AGG | DBC | ^VIX |
| VTI | 1.000 | -0.015 | 0.416 | -0.722 |
| AGG | -0.015 | 1.000 | -0.020 | 0.059 |
| DBC | 0.416 | -0.020 | 1.000 | -0.315 |
| ^VIX | -0.722 | 0.059 | -0.315 | 1.000 |

| | Wilshire Stock | BofA Corp Bond | Fed-funds | Gold |
|---|---|---|---|---|
| Wilshire Stock | 1.000 | -0.130 | -0.007 | 0.010 |
| BofA Corp Bond | -0.130 | 1.000 | 0.003 | 0.054 |
| Fed-funds | -0.007 | 0.003 | 1.000 | 0.013 |
| Gold | 0.010 | 0.054 | 0.013 | 1.000 |

Figure 2 - Correlations

#### 3.2.3 Features experiment

Our third experiment sought to evaluate whether incorporating additional features beyond security price and return, such as financial market and economic indicators, could enhance the results observed in the Oxford study, particularly when extending the time- period to include the pandemic.

Like the first experiment, we used the same architecture, extended time-period, and benchmarking approach.

However, a notable deviation was the addition of 21 new features to the Oxford data. These features comprised two financial market indicators and 19 economic indicators (Figure 1). The data frequency varied across daily, weekly, monthly, and quarterly intervals. To ensure consistency in daily (trading days) data points, we employed forward filling and backfilling techniques. Moreover, we transformed twelve economic indicators reported in units into year-over-year percent changes, ensuring a standardized comparison across the dataset.

### 3.3 Exploring alternative designs and data

We aimed to enhance the Oxford approach by exploring alternative designs and data, addressing potential limitations. This included using a richer historical dataset, integrating state variables for improved regime learning, and exploring LSTM ensemble with state variables and transformer architectures like encoder-only and encoder-decoder models [29]. Building on promising results, we chose the full transformer architecture (Transformer), a sequence-to-sequence model with configurable dimensions and attention mechanisms (Figure 3).

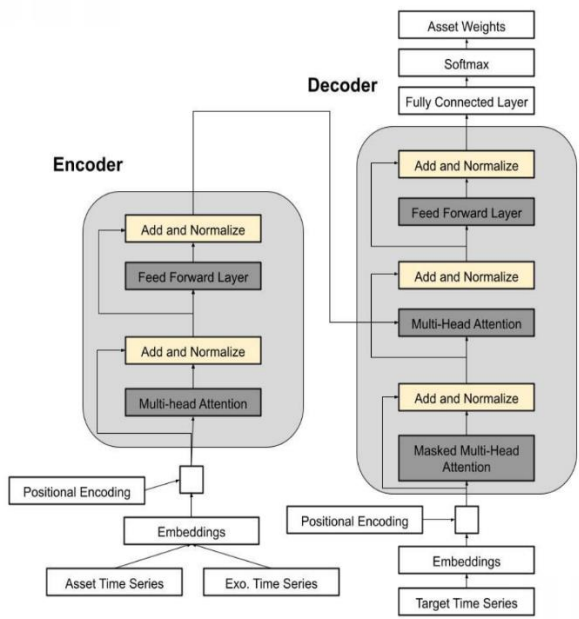

Figure 3 – Transformer architecture

The Oxford study's data limitations led us to hypothesize that pretraining the model on broader historical data from 1986 could improve transfer learning, especially for less represented scenarios in the 2006-2023 period. Consequently, we pretrained the model on a broader dataset with similar asset types, then fine-tuned it on the Oxford data.

For our experiment, we followed a two-phase approach in our experiment: first, we used a pre-training dataset to initialize the model, followed by training on a second dataset spanning the Oxford study's time-period. Although previous experiments showed limited benefits from extra features, recent findings by [30] on slow momentum strategies indicated that adding state features to LSTM learners could improve portfolio Sharpes by enhancing regime identification. This prompted us to expand our feature space to include market and economic data from 1986 onwards. Our pre-training dataset included stock and bond price data from our asset experiments (Wilshire and BofA) dating back to 1986, for initializing weights in VTI and AGG funds. Additionally, we used annualized 30-day historical volatility from Wilshire stock price data to set weights for ^VIX and gold prices for the commodities (DBC) index.

In these experiments, we implemented the Sharpe loss function like the Oxford experiments. To enhance model robustness, we added L2 Regularization to the Adam optimizer and included a dropout layer for normalization and to counter overfitting. Hyperparameter tuning used a 10% validation set from each training period, with testing on a search grid. Final model settings included batch size=128, dropout=0.05, L2 regularization=1e-5, lookback window=504, number of heads=2, layers=1, embedding size=32, learning rate=0.001, and epochs=50. Transformer tests required retraining every 2 years to compute test weights for the subsequent 2 years.

## 4. Results

### 4.1 Replication and verification

During training, we analyzed the training and validation curves of our LSTM model on the Oxford dataset (Figures 4, 5). These curves revealed five distinct training patterns, with most validation curves peaking in Sharpe ratios between 20 and 40 epochs. One curve improved until epoch 100, while another peaked around epoch 90, suggesting the Oxford team might have stopped training at epoch 100 for generalization and to avoid overfitting.

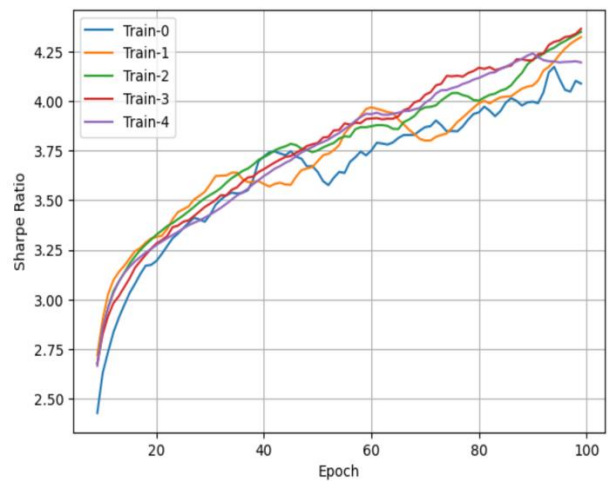

Figure 4 – Training curves

Figure 6 shows the dynamic nature of our LSTM model in reallocating capital across asset classes, with significant variations in allocations to VTI (0.1 - 0.6) and AGG (0.2 - 0.9), while allocations to DBC and ^VIX ranged from 0.0 to 0.07 and 0.0 to 0.25, respectively.

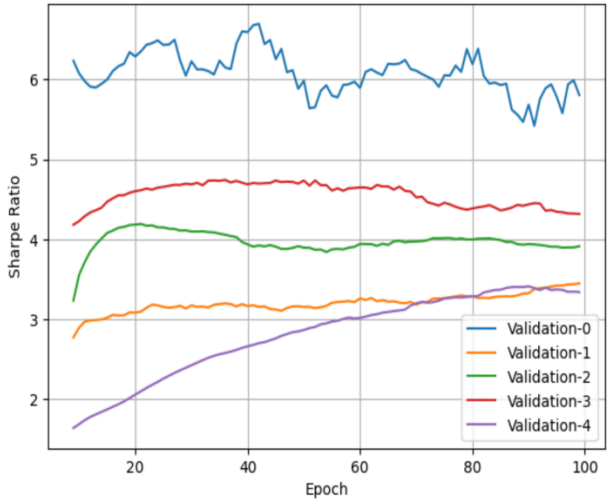

Figure 5 – Validation curves

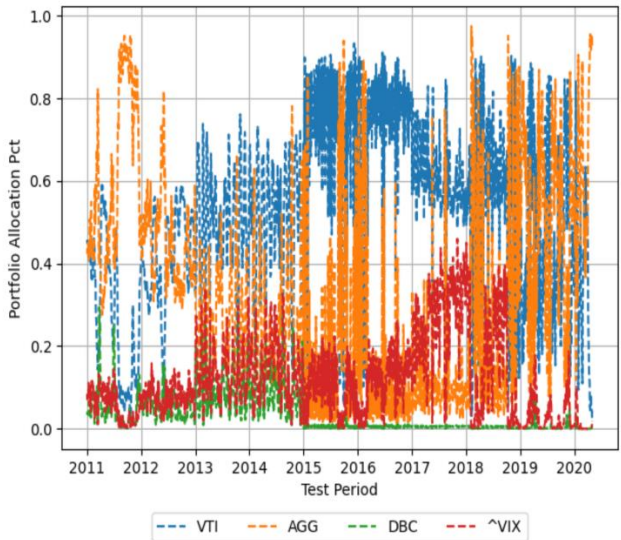

Figure 6 – Dynamic asset weights

Considering the active trading nature of our LSTM model, like the approach used in the Oxford study, we accounted for a 0.01% transaction fee on daily adjustments in allocation weights to determine net returns.

Table 1 demonstrates the performance of our LSTM model (net of transaction costs) against the Oxford results and our benchmark portfolios. Our replicated portfolio's Sharpe ratio closely aligns with the Oxford study (1.828 vs 1.858), demonstrating superior returns in comparison to the MVO and Balanced portfolios in terms of annual return, profitability (percent of positive returns), and risk metrics (max drawdown).

We conducted 30 iterations of LSTM modeling due to its stochastic nature, followed by a two-tailed z-test to compare results with Oxford's. The z-statistic of 1.13, below the critical value of 1.96 at alpha 0.05, indicates no statistically significant difference in mean Sharpe ratio between our LSTM and Oxford's. However, z-tests revealed differences in mean Sharpe values for MVO and Balanced portfolios. Figure 7 shows 1-year rolling Sharpe ratios of our LSTM versus benchmarks, consistently demonstrating LSTM's superiority. MW test results confirmed significantly higher Sharpe ratios for our LSTM model versus benchmarks. Additionally, LSTM outperformed MVO and Balanced portfolios on 72% and 90% of days, respectively, during the test period. Finally, Figure 8 compares return and risk characteristics of the LSTM portfolio to benchmark portfolios and assets, highlighting the LSTM model's superiority once again.

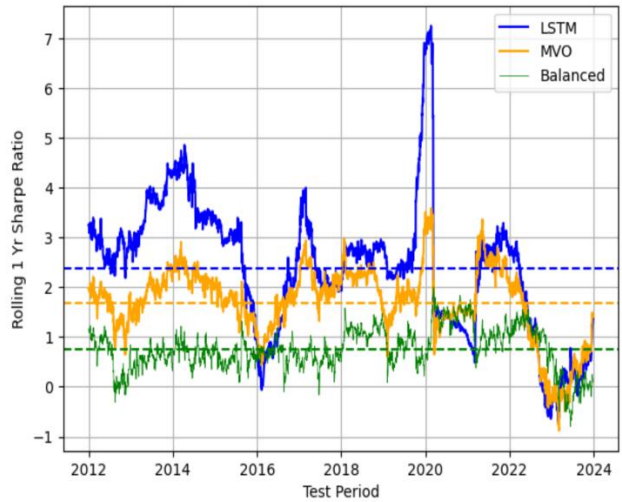

Figure 7 – Verification rolling Sharpe

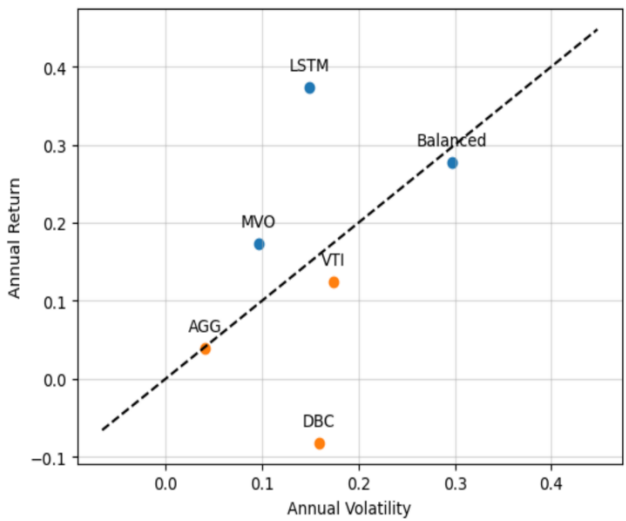

Figure 8 – Verification LSTM risk-return

## 4.2 Time experiment

The authors noted the LSTM's initial strong performance at the start of the COVID-19, but their analysis concluded in April 2020, just as the crisis began. To thoroughly evaluate the LSTM model's abilities in challenging market conditions, we extended our analysis until December 2023, capturing the pandemic's peak volatility.

We maintained the LSTM's hyperparameter settings and expanded the dataset to cover until 12/31/2023. Table 1 illustrates that although the LSTM maintained an overall performance advantage, this edge slightly diminished, as the annual return of the LSTM fell below that of the Balanced portfolio.

During the 5/1/2020 to 12/31/2023 period, Figure 9 indicates a convergence of the LSTM and MVO portfolio's rolling Sharpe ratios, with the LSTM's mean rolling Sharpe ratio at 1.29 compared to 1.26 for MVO. The LSTM's outperformance of MVO decreased from 72% to 48% of days, and the MW test yielded a non-significant p-value of 0.926, indicating no statistically significant difference in their Sharpe ratios. Figure 10 depicts the risk-return performance during the same period, showcasing the LSTM's risk-adjusted return alignment with other instruments and a shift in performance dynamics, particularly highlighting its struggle during the COVID period where it performed no better than MVO.

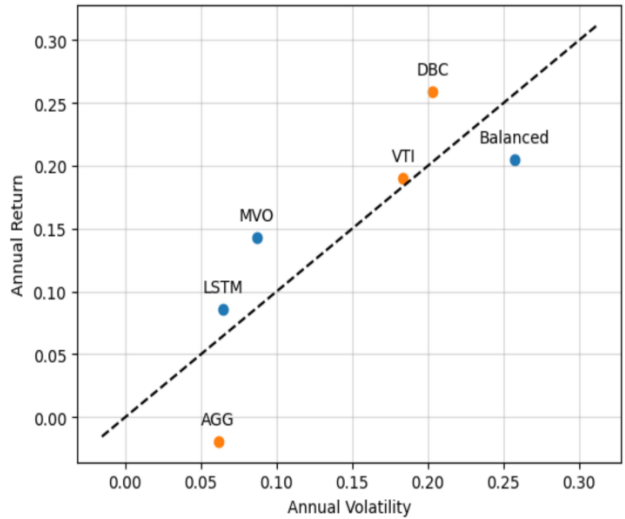

Figure 10 – Time experiment risk-return

## 4.3 Asset experiment

The second experiment aimed to assess the applicability of the Oxford LSTM methodology across diverse asset classes, prioritizing asset categories with minimal or negative correlations.

We trained the LSTM model using four additional assets and extending the analysis period until 12/31/2023. While experimenting with various hyperparameter configurations, we ultimately determined that the original settings yielded the most robust performance on the validation dataset.

Table 1 illustrates the LSTM model's weaker performance compared to the MVO and Balanced portfolios, with a Sharpe ratio of 0.81 versus 1.08 and 0.89, respectively. This indicates lower risk-adjusted returns for the LSTM. It also had a higher drawdown measure of 0.204 compared to 0.059 and 0.142 for the other portfolios. Across the testing period, the LSTM's rolling Sharpe ratios averaged 1.06, slightly below MVO's 1.11 (Figure 11). The MW test found no statistical significance at the 0.05 level between LSTM and MVO, with LSTM outperforming MVO in rolling Sharpe ratios only 0.45 of the time overall and 0.21 post-COVID-19. Post-04/30/2020, LSTM had a mean rolling Sharpe ratio of 0.43, contrasting with 0.91 and 0.72 for MVO and Balanced portfolios, respectively. MW tests indicated statistically higher Sharpe ratios for both benchmark portfolios compared to LSTM (Figure 12).

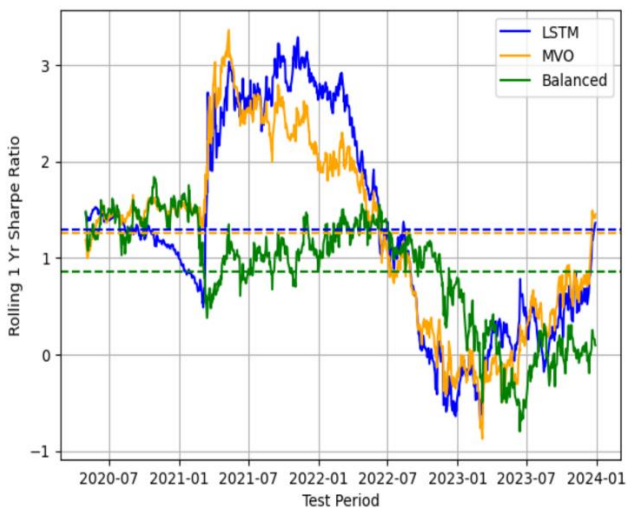

Figure 9 – Time experiment Sharpe

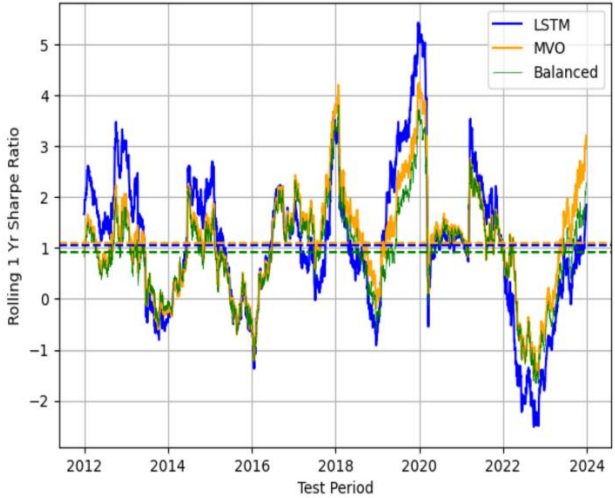

Figure 11 – Asset experiment Sharpe

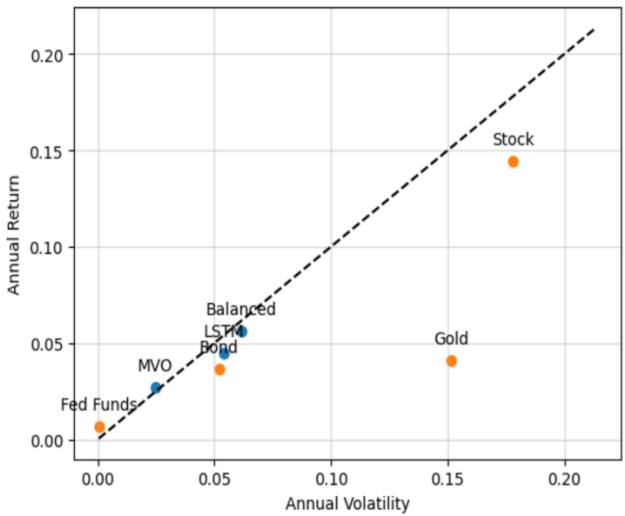

Figure 12 – Asset experiment risk-return

### 4.4 Features experiment

Given the LSTM model's underperformance compared to benchmarks during the pandemic, we sought to enhance its capabilities by expanding the feature space to include financial market and economic indicators. This expansion aimed to uncover intricate relationships, potentially leading to improved predictive accuracy and better generalization.

For training and testing, we utilized the same extended time frame as in previous experiments. Our model was trained on an expanded feature set comprising 29 variables. Notably, we found that the model's performance was optimized with similar hyperparameter configurations as in previous experiments.

While our model exhibited strong performance across the entire analysis period, focusing on the pandemic era revealed notable differences. Figure 13 illustrates that the LSTM's rolling Sharpe ratio was below that of the MVO benchmark approximately 61.5% of the time. Specifically, the mean rolling Sharpe ratio for the LSTM model was 0.62, contrasting with 1.26 for MVO and 0.85 for Balanced. Moreover, a MW test yielded a p-value close to zero, indicating statistical significance at the 0.05 level and suggesting that the LSTM's Sharpe ratio was notably lower than that of MVO during the pandemic. Finally, the risk-return chart in Figure 14 underscores the model's subpar risk-adjusted performance vis-a-vis benchmarks during the pandemic period.

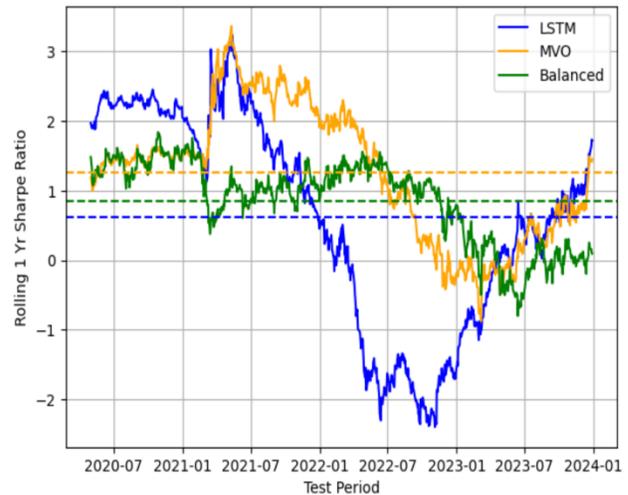

Figure 13 – Features experiment Sharpe

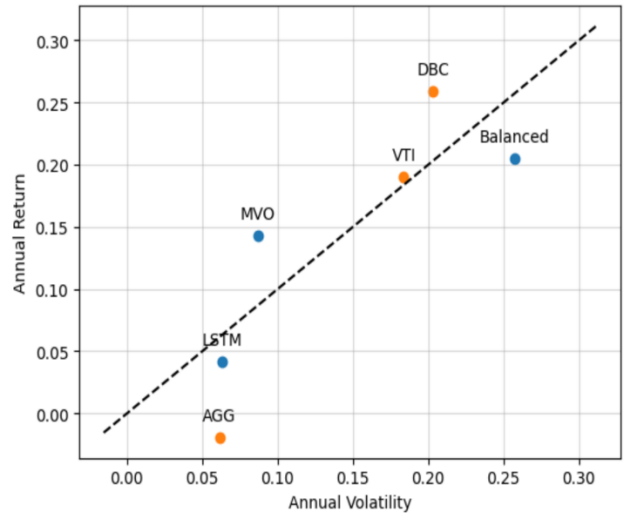

Figure 14 – Features experiment risk-return

### 4.5 Alternative designs and data

Our final experiment aimed to evaluate whether pre-training, expanded features, and the transformer architecture could outperform the Oxford LSTM model. Table 1 presents metrics for the Transformer and LSTM models, alongside MVO and Balanced portfolios. The Transformer achieved a Sharpe ratio of 2.7, surpassing others in risk metrics, positive returns, and achieving an annual return like MVO. Statistical analysis confirmed the Transformer's superiority, with significantly higher rolling Sharpe ratios compared to benchmarks; the MW p-value was close to zero for all three comparisons, indicating that the Transformer's Sharpe was statistically higher than the other benchmarks (Figures 15, 16). It consistently outperformed LSTM, MVO, and Balanced portfolios, demonstrating robustness and effectiveness in portfolio optimization. Specifically, the Transformer model produced higher rolling Sharpe ratios on 0.72, 0.90, and 0.92 of the days over the entire test period versus LSTM, MVO, and Balanced, respectively, and 0.68, 0.69, and 0.74, respectively, of the days during the zoom-in period post 05/01/2020. Figure 17 illustrates the superior risk vs. return profile of the Transformer compared to benchmarks.

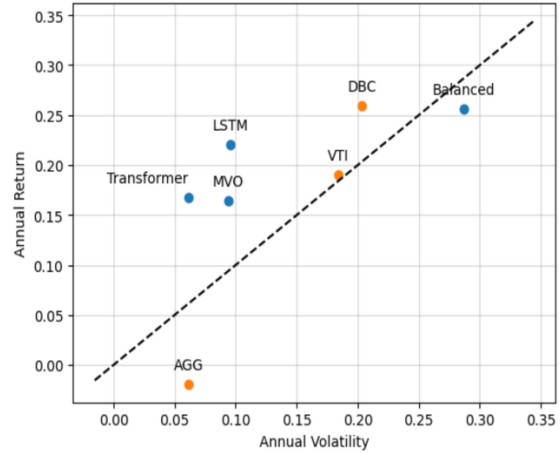

Figure 17 – Transformer risk-return

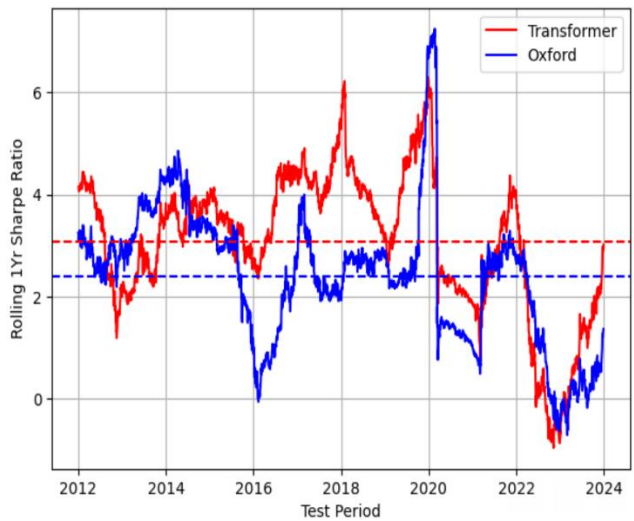

Figure 15 – Transformer Sharpe

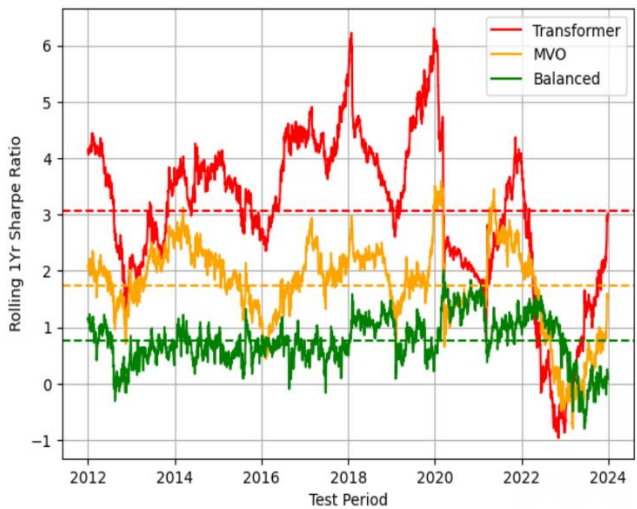

Figure 16 – Transformer vs benchmarks Sharpe

## 5. Discussion

Our interest in DL portfolio optimization stemmed from the Oxford study, which caught our attention for its potential robustness in long-only, multi-asset integrated optimization and its ability to mitigate overfitting common in portfolio optimization. Our objective was to rigorously test the reliability and transferability of their approach. Our time experiment highlighted the resilience of the LSTM model in volatile markets, although it didn't show statistical superiority over traditional methods. Similarly, the asset experiment revealed challenges in adapting to new assets post-COVID-19, impacting performance against benchmarks. Furthermore, our features experiment emphasized the necessity for more sophisticated modeling approaches.

Recognizing room for improvement in the Oxford model, we introduced a Transformer model that exhibited superior adaptability and performance, especially with transfer learning data. This underscores the significance of advanced ML architectures for navigating dynamic market conditions and suggests avenues for enhancing predictive accuracy and adaptability in portfolio optimization strategies. These insights pave the way for future advancements in ML applications for financial management.

## 6. Conclusion

Our exploration of DL portfolio optimization reveals the crucial challenge of navigating market cycles effectively. Long-term performance in such portfolios hinges on understanding and adapting to extended price trends. Addressing key challenges involves training models with limited regime data, possibly through pre-training methods, and developing models that can capture state variables for precise regime recognition, such as transformer architectures. These challenges underscore the ongoing evolution and complexity of DL-driven portfolio optimization, guiding future research toward enhancing predictive accuracy and optimizing strategies under dynamic market conditions.